\documentclass[aps,twocolumn,superscriptaddress,showpacs, nofootinbib, amsmath, amssymb]{revtex4}
\usepackage{graphicx} 
\usepackage{wrapfig}
\usepackage{bbding}
\usepackage{color}
\usepackage{bm}
\usepackage{natbib}
\usepackage{float}
\usepackage{multirow}
\usepackage{array}
\usepackage{physics}
\usepackage{siunitx}
\usepackage[colorlinks=true, urlcolor=blue, citecolor=black, linkcolor=black, pdfborder={0 0 0}]{hyperref}

\newcommand{\PCP}{PrCd$_3$P$_3$}
\newcommand{\NCP}{NdCd$_3$P$_3$}

\newcommand{\wn}{cm$^{-1}$}

\begin{document}

\title{Raman scattering spectroscopic observation of a ferroelastic crossover in bond-frustrated \PCP}
\author{Jackson Davis}
\affiliation{Department of Physics and Astronomy, Johns Hopkins University, Baltimore, Maryland 21218, USA}
\author{Jesse Liebman}
\affiliation{Department of Physics and Astronomy, Johns Hopkins University, Baltimore, Maryland 21218, USA}
\author{Dibyata Rout}
\affiliation{Materials Department, University of California, Santa Barbara, California 93106, USA}
\author{S.J. Gomez Alvarado}
\affiliation{Materials Department, University of California, Santa Barbara, California 93106, USA}
\author{Stephen D. Wilson}
\affiliation{Materials Department, University of California, Santa Barbara, California 93106, USA}
\author{Natalia Drichko}
\affiliation{Department of Physics and Astronomy, Johns Hopkins University, Baltimore, Maryland 21218, USA}
\date{\today}

\begin{abstract}
 2D magnetism in triangular lattices has already shown potential for hosting exotic magnetic states. Control of these magnetic states, both in terms of magnetic properties and in terms of charge doping would be the next step. This makes materials which combine triangular lattice magnetic layers with layers hosting interesting structural or electronic properties particularly useful. \PCP, studied in this work, is one of a family of materials where triangular lattice layers of magnetic rare earth ions alternate with semiconducting hexagonal CdP layers. Using Raman scattering spectroscopy we uncover a structural instability in the CdP layers, associated with a soft mode behavior of a phonon in these layers.   Raman scattering detects crystal electric field excitations, and confirms a singlet ground state for Pr$^{3+}$ and splitting of the doublet levels as a result of the structural instability in CdP layers. While  Pr$^{3+}$ is non-magnetic in \PCP\, we speculate that this family of materials can realize control of the magnetic layer through the CdP layer which can become ferroelectric under strain that would relieve frustration. 
\end{abstract}

\maketitle

\section{Introduction}

In condensed matter physics and materials science,  so-called emergent phases, the states of matter which originate from strong interactions between electrons or their spins, are one of the main areas of research. Strong interactions for magnetic degrees of freedom lead to  ordered magnetic phases, while strong coupling on a geometrically frustrated lattice is known to suppress order, resulting in ordered exotic magnetic ground states~\cite{Starykh2015}, or disordered ones, such as the resonating valence bond state, first suggested to host a quantum spin liquid~\cite{Balents2010,Broholm2020}. 
For electronic degrees of freedom, strong interactions can  result in insulating charge ordered and Mott insulating phases, as well as exotic behavior close to the phase transition~\cite{Imada1998}.  Similarly,  for the charge degrees of freedom, geometrical frustration can prevent a phase transition into an ordered state and  result in exotic charge states~\cite{Hotta2010,Subires2025}. 

Lots of progress was made in terms of understanding the origin of these phases, and now we strive to control them.  Such control could be extrinsic, with, for example, strain or magnetic field, or intrinsic, where we combine properties such that one degree of freedom determines the behavior of the other. Such an approach is successfully developing in, for example, intercalated two-dimensional materials, resulting in magnetic Weyl semimetals~\cite{Inoshita2019,Ray2025}. Another more traditional example of such combinations is multiferroicity, which combines magnetic and ferroelectric orders, typically on different sublattices of the same material~\cite{Fiebig2016}. 

The combination of magnetism, exotic charge properties, and geometric frustration in one material is of high interest and is demonstrated in the family of materials $\textit{Ln}\textit{M}_3\textit{Pn}_3$ (\textit{Ln} = trivalent lanthanide ion, \textit{M} = Zn or Cd, \textit{Pn} = P or As). In this material, magnetism comes from the triangular lattice of \textit{Ln} atoms in one layer and has been shown to be highly frustrated~\cite{Chamorro2025}, and the interstitial hexagonal layer is shown to produce a short range structural order which leads to formation of Cd-P pairs~\cite{GomezAlvarado2025}. The hexagonal layer can be doped, providing a means to place a conducting electron system in the vicinity of the frustrated antiferromagnetic triangular lattice.~\cite{GomezAlvarado2025}. \PCP, studied in this work, belongs to this family, with the potentially magnetic layer formed by a triangular lattice of  Pr$^{3+}$ ions with $J = 4$. As suggested in Ref.~\cite{GomezAlvarado2025} and confirmed in this  work, in \PCP\ Pr$^{3+}$ has a singlet ground state due to crystal field splitting of the $J = 4$ multiplet, and thus is nonmagnetic. \PCP\ provides an opportunity to study the effect of the structural instability in the hexagonal CdP layer on the local environment of the rare earth ion via the splitting of crystal field non-Kramers doublets. 

In this work we use Raman scattering spectroscopy to probe how lattice and a rare earth ion crystal electric field (CEF) degrees of freedom,  which belong to different layers, are interacting in this material. As typical for rare earth, Pr$^{3+}$ has  highly localized and anisotropic 4$f$ electrons, with energy scales dependent on the crystal field-determined splitting of the atomic orbitals, which can be determined by Raman spectroscopy. The results of temperature-dependent polarized Raman spectroscopy on \PCP\ allows us not only assign CEF levels and phonons, but also identify a displacive structural instability at temperatures around 70~K occurring primarily in the adjacent non-magnetic Cd-P hexagonal layer, and influencing the symmetry and crystal field levels  of the Pr-based triangular lattice.

\begin{figure}[h]
    \centering
    \includegraphics[width=0.9\linewidth]{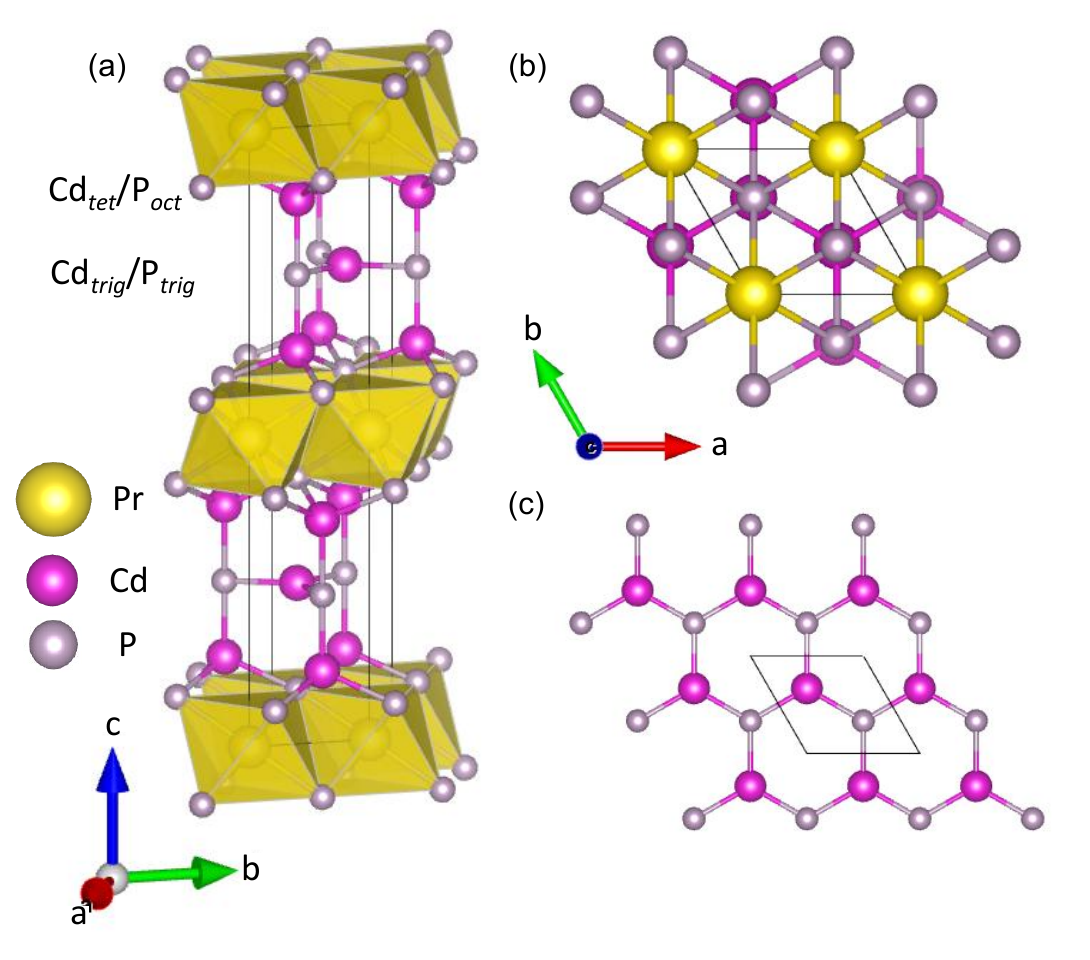}
    \caption{Structure of \PCP, viewed along (a) skewed crystallographic a-axis and (b) [001] direction. (c) Cd$_{trig}$/P$_{trig}$ layer viewed along the [001] direction.}
    \label{fig:Fig_structure}
\end{figure}

\section{Experimental}

While \PCP\  was a primary interest of this work, we used measurements on \NCP\ to confirm the identification of  phonons and CEF. Single crystals of \PCP\ and \NCP\ were synthesized by a molten salt flux method as previously reported \cite{Nientiedt1999}.

Raman scattering spectra were measured for single crystals of \PCP\ and \NCP\ of $\sim$ 0.7 x 0.7 mm$^2$ size in pseudo-Brewster angle scattering geometry using a Horiba Jobin-Yvon T64000 Spectrometer. Raman scattering was excited with the 514.5 nm line of an Ar$^+$/Kr$^+$ ion laser.  The excitation beam was  focused on a spot size of approximately 50 x 100 $\mu$m$^2$.  Spectra were recorded  with a spectral resolution of  4 \wn (0.5 meV).  For low temperature measurements samples were mounted on a cold finger of a Janis ST-500 He flow cryostat using GE varnish. Spectra were measured using laser power of 10 mW for higher temperatures, and  3 mW at the lowest temperature. The laser heating of 1~K/1 mW  in the used setup is estimated based on the comparison of Stokes and anti-Stokes Raman spectra for this and previously studied materials.

Raman scattering was measured in the $(ab)$ crystallographic plane of  \PCP\ and \NCP\ single crystals. The crystals were oriented using linearly polarized Raman scattering, using symmetry selection rules listed in Table~\ref{tab:pol_table}.~
Spectra of \PCP\ were additionally measured in circular polarization RR and RL, where circular polarization of light was defined relative to the reference frame of the sample, in the spectral range from 1.7 to 83.4 meV. Due to the low intensity of the Raman response, in order to reach frequencies down to 0.6 meV for the soft mode detection, the spectra were measured with the polarization of excitation light $e \parallel a$, where $a$ is a crystallographic axis, and without an analyzer. 
The  intensity  of the Raman response of both materials was extremely low at 300 K, with the strongest features having intensities below 1.5 counts/sec, due to the semiconducting properties of the materials. In order to reduce the signal-to-noise ratio, we used accumulations of 60 min per spectrum, where the accumulation time is limited by the mechanical stability of the system.


Both \PCP\ and \NCP\ crystallize in space group \textit{P}$6_3$/\textit{mmc} with corresponding point group \textit{$D_{6h}$} \cite{GomezAlvarado2025}. Table~\ref{tab:wyckoff} displays the $\Gamma$-point phonons associated with the Wyckoff positions. Symmetry constraints dictate a total of eight (2A$_{1g}$ + 2E$_{1g}$ + 4E$_{2g}$) Raman-active phonons arising from the two crystallographically distinct Cd and P sites in \PCP\ and \NCP, with A$_{1g}$ and E$_{2g}$  active in the scattering from the $(ab)$ plane. Their polarization dependence is summarized in Table~\ref{tab:pol_table}. 


\begin{table}[h]
    \centering
    \begin{tabular}{|c c c|}
         \hline
         Element & Wyckoff position & Phonon symmetry \\
         \hline
         Pr & 2a & inactive \\
         Cd$_{tet}$ & 4f & A$_{1g}$ + E$_{1g}$ + E$_{2g}$ \\
         Cd$_{trig}$ & 2d & E$_{2g}$ \\
         P$_{oct}$ & 4f & A$_{1g}$ + E$_{1g}$ + E$_{2g}$ \\
         P$_{trig}$ & 2c & E$_{2g}$ \\
         \hline
    \end{tabular} 
    \caption{Wyckoff positions and symmetry representations of related $\Gamma$- point phonons  in \PCP, following naming convention for Cd$_{trig}$/P$_{trig}$ in Ref. \cite{GomezAlvarado2025}.}
    \label{tab:wyckoff}
\end{table}

\begin{table}[h]
    \centering
    \begin{tabular}{|c c|}
        \hline
        Scattering geometry & Active symmetry species \\
        \hline
        XX & A$_{1g}$ + E$_{2g}$\\
        YY & A$_{1g}$ + E$_{2g}$\\
        XY & E$_{2g}$\\
        RR & A$_{1g}$\\
        RL & E$_{2g}$\\
        \hline
    \end{tabular}
    \caption{Polarization dependence of Raman-active phonons in the $(ab)$ plane.}
    \label{tab:pol_table}
\end{table}

\textit{Raman scattering data analysis:} A temperature-independent background was subtracted from all Raman spectra to highlight the behavior of narrow phonon and CEF features. The background is comprised of two components: stray laser light scattering, which can be modeled by a Gaussian profile centered at 0 \wn,  and a  background related to photoluminescence, the maximum of which is expected at frequencies above the measured spectral range,  approximated by a straight line in the 0-500 \wn\ spectral region (see SI for spectra before subtraction). 


Temperature dependence of the parameters of phonons \textbf{Ph2-Ph6} was obtained by a least-squares fit of each phonon to a Gaussian peak. Due to the low Raman intensity, high-resolution/low-intensity measurements were not feasible, and the Gaussian profile of the phonons reflects this resolution limitation rather than an intrinsic disorder in the sample. However, \textbf{Ph1}, which is broader than the other phonons at low temperatures, was fit by a Lorentzian profile.


We perform ab initio calculations of the electronic band structure and vibrational modes using the open source density functional theory software \textsc{Quantum ESPRESSO}~\cite{Giannozzi_2009, Giannozzi_2017}. 
For electronic band structure calculations, we use an $8 \times 8\times 8$ Mokhorst-Pack grid, a plane-wave energy cutoff of 100~Ry, and a convergence threshold of $10^{-18}$~Ry. 
For Pr, we used the pseudopotential Pr.pbe-spdn-rrkjus\_psl.1.0.0.UPF from \texttt{quantum-espresso.org}~\cite{DalCorso2014}, and for Cd and P we use the psuedopotentials cd\_pbesol\_v1.uspp.F.UPF and p\_pbesol\_v1.5.uspp.F.UPF from the GBRV pseudopotential library~\cite{Garrity2014}, using the generalized gradient approximation~\cite{Perdew1996} for all pseudopotentials. 
For structural relaxation calculations we used a total energy convergence threshold of $10^{-8}$~Ry and force threshold of $10^{-6}$~Ry/Bohr. 
We calculate vibrational frequencies at the $\Gamma$-point using density functional perturbation theory, using acoustic sum rules. 
We use the frequency and irreducible representations of each calculated mode to compare to the observed modes in Raman scattering spectra.

\section{Results}



 


The structure of \PCP\ consists of layers of tilted edge-sharing PrP$_6$  octahedra alternating with honeycomb Cd-P layers, with Cd and P occupying each distinct site of the honeycomb (Fig. \ref{fig:Fig_structure}c). According to the XRD measured at 300 K, the unit cell has \textit{P}$6_3$/\textit{mmc} symmetry.

Using a combination of DFT phonon calculations, a comparison to the low temperature spectra of \NCP, and a comparison to inelastic neutron scattering spectra of \PCP\ powder~\cite{GomezAlvarado2025}, we assign the excitations in the Raman scattering spectra to phonons or crystal electric field (CEF) transitions accordingly.


\begin{figure*}[t]
    \centering
    \includegraphics[width=\linewidth]{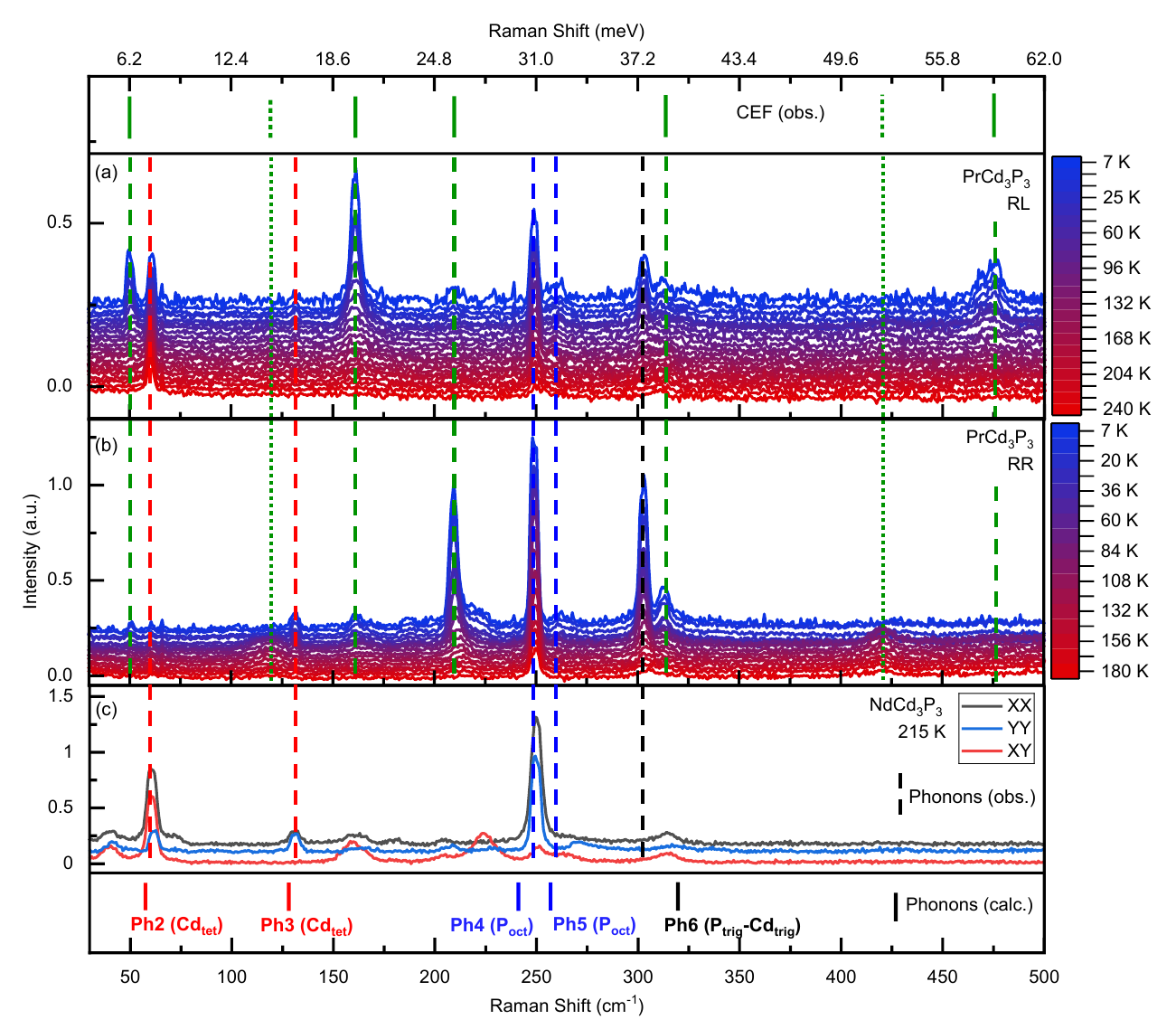}
    \caption{Raman scattering spectra of \PCP\ with  frequencies of Raman active phonons  and crystal field excitations marked by vertical lines. (a) spectra in RL scattering channel between 7 and 240~K, (b) spectra in RR scattering channel between 7 and 180~K. (c) Raman scattering  spectra of \NCP\ in XX, YY, and XY scattering channels at 215 K. Black, red, and blue vertical dashed lines show observed frequencies of phonons of \PCP, calculated frequencies are marked in the bottom panel, see Table~\ref{tab:phonon_table} for exact frequencies. Green vertical dashed lines show observed positions of crystal field excitations, with the dashed lines with shorter dashes marking observed interband crystal field excitations (see Table~\ref{tab:CEFcalc_table} for exact frequencies). The comparison of \PCP and \NCP\ spectra allows one to identify phonons vs crystal field excitations of Pr$^{3+}$ in the Raman scattering spectra of \PCP.  }
    \label{fig:Fig_warming_Raman}
\end{figure*}

\subsection{Phonon spectra}


\begin{figure*}[ht]
    \centering
    \includegraphics[width=0.95\linewidth]{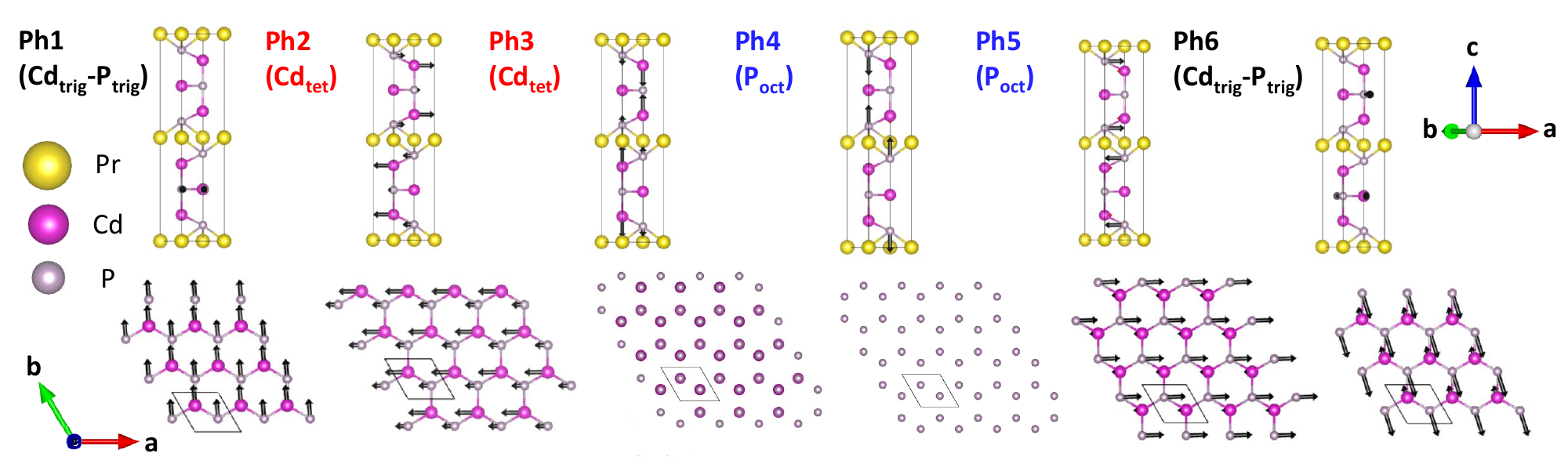}
    \caption{$\Gamma$-point atomic displacements for six DFT-calculated phonons of \PCP. Phonons are labeled by \textbf{Ph1}-\textbf{Ph6}, with the dominant atomic motion of each mode noted in brackets. The phonon labels are color-coded dependent on which layer the  atoms belong to. Top: atomic displacements viewed along [120] direction. Bottom: atomic displacements viewed along [001] direction for single layers of displaced atomic sites.}
    \label{fig:phonons}
\end{figure*}

Fig.~\ref{fig:Fig_warming_Raman}a(b) presents Raman scattering spectra of \PCP$ $ collected in RL(RR) polarization from 7 K to 240 K(180 K) in the spectral range between 3.7 and 62.0 meV. 
~For comparison, spectra of isostructural NdCd$_3$P$_3$ measured at 215 K are shown in Fig.~\ref{fig:Fig_warming_Raman}c. The temperature dependence of unpolarized spectra in the lowest frequency range, 0.5 to 2.5 meV, 
~down to 13.5 K is presented in Fig. \ref{fig:Fig_softmode}. Results of the DFT phonon calculations  performed assuming  \textit{P}$6_3$/\textit{mmc} symmetry of the unit cell corresponding to the high-temperature crystal structure  are summarized in Table~\ref{tab:phonon_table}, and $\Gamma$-point atomic displacements of each mode are shown in Fig. \ref{fig:phonons}. 

Based on the symmetry analysis  assuming  \textit{P}$6_3$/\textit{mmc} symmetry of the unit cell, we expect 6 Raman-active phonons (2A$_{1g}$+4E$_g$) from scattering in the $(ab)$ plane for both \PCP\ and \NCP, shown in Table~\ref{tab:wyckoff}.  
 Room temperature spectra of \PCP\ have very low intensity due to the semiconducting properties of this material; In these spectra  we observe one of the two expected A$_{1g}$ phonons  (30.9 meV) and three of the four expected  E$_{2g}$ phonons (1.5, 7.6, and 37.6 meV), see Table~\ref{tab:phonon_table}.  Twelve excitations, which appear in the spectra of  \PCP\  at low temperatures,  can originate from phonons or CEF excitations. A comparison to NdCd$_3$P$_3$, which is expected to have similar phonon frequencies but different CEF energies, and to the calculated phonon frequencies (Table~\ref{tab:phonon_table})  assists the assignment. Based on this comparison we assign the mode observed at 16.3 meV to  an A$_{1g}$ phonon, and the mode at 32.3 meV to an E$_{2g}$ phonon.




\begin{table}[h]
    \centering
    \begin{tabular}{|c|c|c|c|}
         \hline
         Phonon & \multicolumn{3}{c|}{Frequency (meV)} \\
         & \PCP (calc) & \PCP (obs) & \NCP (obs) \\
         \hline
         \textbf{Ph1} (E$_{2g}$) & 3.4 & 1.5 (XX+XY) & 1.8\\
         \textbf{Ph2} (E$_{2g}$) & 7.1 & 7.6 (RL) & 7.6\\
         \textbf{Ph3} (A$_{1g}$) & 15.9 & 16.3 (RR) & 16.2\\
         \textbf{Ph4} (A$_{1g}$) & 29.9 & 30.9 (RR) & 31.0\\
         \textbf{Ph5} (E$_{2g}$) & 31.9 & 32.3 (RR+RL) & 32.3\\
         \textbf{Ph6} (E$_{2g}$) & 39.6 & 37.6 (RR+RL*) & 38.9\\
         \hline
    \end{tabular}
    \caption{Observed and calculated frequencies of phonons in \PCP$ $ (meV). Dominant polarizations for each phonon are noted next to observed frequencies. *For \textbf{Ph6}, polarizations at room temperature are used. Observed phonon frequencies in NdCd$_3$P$_3$ are presented for comparison.}
    \label{tab:phonon_table}
\end{table}


\subsection{Temperature Dependence of Phonons}

\begin{figure}[h]
    \centering
    \includegraphics[width=\linewidth]{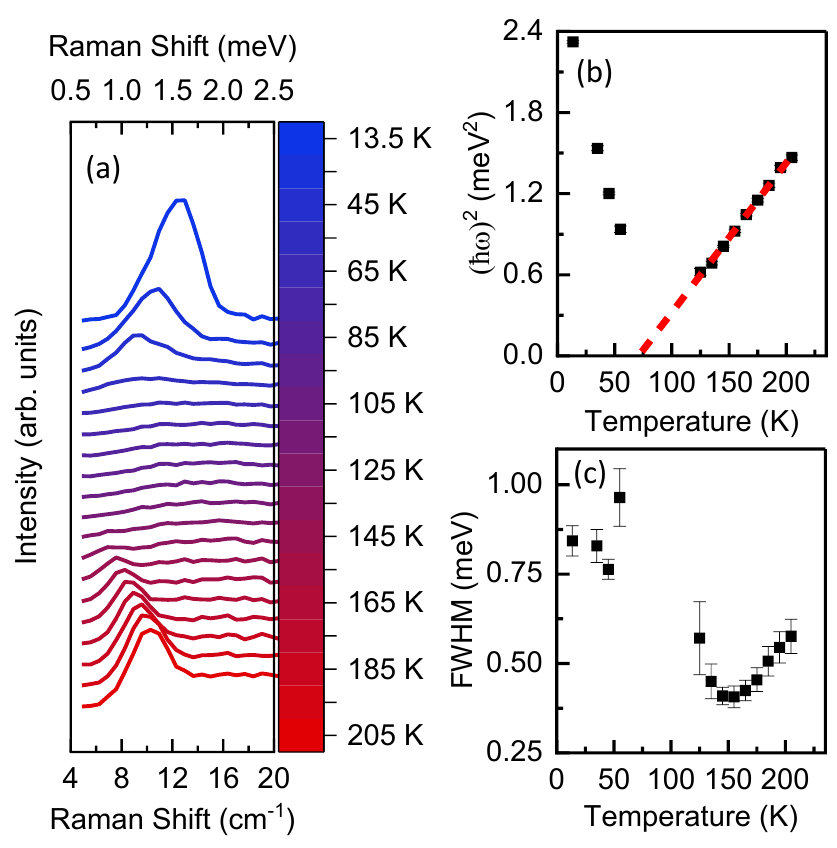}
    \caption{Temperature dependence of the lowest frequency $E_{2g}$ phonon (Ph1). (a) Raman spectra in the spectral region around the phonon, the intensity is  normalized to spectral weight of A$_{1g}$ phonon at 30.8 meV. (b) Temperature dependence of the squared soft mode energy (black squares), showing softening and subsequent hardening in the low temperature regime. A  fit to a linear model in the high-temperature regime is shown by red dashed line and yields T=70~K for the structural transition temperature. (c) Temperature dependence of the full-width at half maximum (FWHM) of the Lorenztian phonon lineshape.}
    \label{fig:Fig_softmode}
\end{figure}

\begin{figure}
    \centering
    \includegraphics[width=\linewidth]{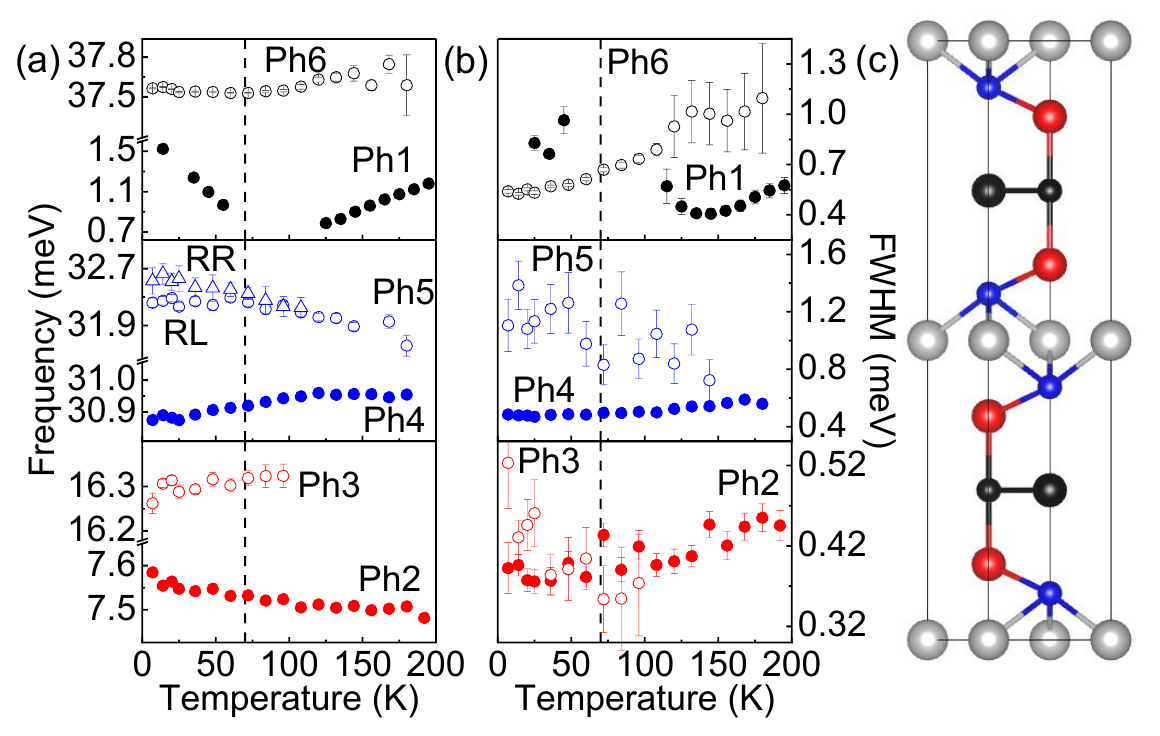}
    \caption{Temperature dependence of (a) phonon frequencies and  (b) width (FWHM); The phonon labels are color-coded dependent on which layer the  atoms belong to, as marked in the \PCP\ crystal structure shown in panel (c). Top: phonons involving motion of the interstitial layer. Middle: phonons involving motion of the P$_{oct}$ layer. Bottom: phonons involving motion of the Cd$_{tet}$ layer. The structural instability temperature obtained from the soft mode (70 K) is marked with a dashed vertical line.}
    \label{fig:phonon_Tdep}
\end{figure}

In the spectra in Fig~\ref{fig:Fig_warming_Raman}, one can easily notice an appearance of modes at low temperatures. In many cases these excitations become observable due to the decrease of the linewidth at low temperatures.  Below we discuss the temperature dependence of the phonons, organizing them by their structural layer, which correlates with the magnitude and nature of changes on cooling.


{\it Phonons of Cd$_{trig}$/P$_{trig}$ honeycomb layer:} 
The behavior of the phonons \textbf{Ph1} (1.5 meV) and \textbf{Ph6} (37.6 meV) corresponding to displacements of the Cd$_{trig}$/P$_{trig}$ sites in the interstitial CdP honeycomb layer (Fig. \ref{fig:phonons}) appear to be the first step to an understanding of the low temperature state of \PCP. 
Fig.~\ref{fig:Fig_softmode}a shows the temperature dependence of E$_{2g}$ \textbf{Ph1}.  This mode involves  the in-plane motion of the Cd$_{trig}$/P$_{trig}$ sites, where Cd$_{trig}$ and P$_{trig}$ sites move with different amplitudes along Cd-P bonds,  resulting in distorted hexagons (Fig.~\ref{fig:phonons}).  The mode shifts down  from 1.3 to 0.8 meV upon cooling from 205 K to 125 K, then moves below the accessible spectral range. Below 55 K, the phonon reappears at 1 meV and shifts up to  1.6 meV at 13.5 K. This is a characteristic behavior of a soft mode \cite{Dove1997}, which can mediate a structural transition or crossover.  As expected for a soft mode, the square of the phonon energy above the transition follows a linear $T$-dependence, with $(\hbar\omega)^2 \propto (T-T_c)$\cite{Dove1997}. A linear fit (Fig.~\ref{fig:Fig_softmode}b) yields a transition temperature of approximately $T_c = 70$ K. The width of the excitation diverges when the phonon softens (Fig.~\ref{fig:Fig_softmode}c). When the mode emerges in the accessible frequency range below 55~K, its width remains larger than at higher temperatures, pointing to the relieved degeneracy of this E$_g$ mode in the low temperature state. 

The higher-energy  Cd$_{trig}$/P$_{trig}$ E$_{2g}$ phonon \textbf{Ph6} is dominated by in-plane P$_{trig}$ motion (Fig.~\ref{fig:phonons}) and is expected at 39.6 meV according to the calculations. Of the excitations observed in the relevant range, the one found at 37.6 meV at low temperatures in both  RR and RL is a good candidate. The intensity of this phonon increases considerably on cooling, and the frequency softens, reaching a constant 37.6 meV below $\sim$120 K. (Fig.~\ref{fig:phonon_Tdep}a). It narrows on cooling and reaches a constant width of 0.6 meV below 40 K, following the temperature dependence predicted by the Klemens model of phonon decay (Fig. \ref{fig:phonon_Tdep}b)~\cite{Klemens1966}.  Despite the decrease of the linewidth on cooling, it is wider than all other observed phonons, except for the weak \textbf{Ph5}. For a doubly degenerate phonon this suggests splitting which is not resolved due to its small amplitude. Another excitation observed in this range  at  38.5 meV in RL is assigned to a CEF excitation, as discussed below. This CEF excitation does not vibronically couple to the phonon, since the phonon eigenvector does not involve motion of the atoms from Pr$^{3+}$ environment.


{\it Phonons of the Cd$_{tet}$ layer:} Cd atoms of this layer are found in between the Cd$_{trig}$/P$_{trig}$ hexagonal layer, where we observe the structural transition, and a layer of Pr$^{3+}$ centered octahedra. ~ \textbf{Ph2} (E$_{2g}$ at 7.6 meV) and \textbf{Ph3} (A$_{1g}$ at 16.3 meV) phonons (marked red in Fig.~\ref{fig:Fig_warming_Raman}, \ref{fig:phonons}, \ref{fig:phonon_Tdep}) correspond  to displacements of these Cd$_{tet}$ sites  whose eigenvectors are shown in Fig. \ref{fig:phonons}. \textbf{Ph2}, dominated by motion of the Cd$_{tet}$ ions in the (001) plane, hardens by $\sim$0.1 meV and  narrows below the instrumental resolution of 0.5 meV upon cooling. \textbf{Ph3}, associated with out-of-plane Cd$_{tet}$ motion, has very low intensity. It is extremely broad at high temperatures, and is resolved as a coherent excitation only below $\sim$120 K. Below 100 K, no significant changes in the frequency are resolved within the instrumental precision, but a slight broadening is observed at the lowest temperatures.

{\it Phonons of the $P_{oct}$ layer formed by PrP$_6$ octahedra:} Finally, we examine phonons \textbf{Ph4} (A$_{1g}$ at 30.9 meV) and \textbf{Ph5} (E$_{2g}$ at 32.3 meV), marked in blue in Fig.~\ref{fig:Fig_warming_Raman}, \ref{fig:phonons}, and \ref{fig:phonon_Tdep}, corresponding to displacements of the P$_{oct}$ sites which comprise the octahedral environment of Pr (Fig. \ref{fig:Fig_structure}). \textbf{Ph4}, associated with out-of-plane motion of P$_{oct}$ (Fig.~\ref{fig:phonons}), is observed at frequencies higher than calculated (29.9 meV). It is intense in both polarizations, but stronger in RR. It shows continuous softening upon cooling below 120~K, while the behavior of the linewidth is conventional (see Fig.~\ref{fig:phonon_Tdep}). \textbf{Ph5}, associated with in-plane P$_{oct}$ motion (Fig.~\ref{fig:phonons}) has very low intensity in the whole measured temperature range, and is broad at all temperatures. Below 70~K frequencies associated with this phonon differ by  $\sim0.2$ meV between RR and RL, suggesting a splitting of this E$_{2g}$ phonon.

Overall, temperature dependence of the phonons in \PCP\ suggests a structural phase transition at around 70~K, associated with a soft mode in the P-Cd trigonal layer.   Following previous XRD studies \cite{GomezAlvarado2025}, we can expect a symmetry breaking leading to D$_{2h}$ symmetry of the unit cell at low temperatures.  However, the splitting of the E$_{2g}$ phonon modes expected at  such symmetry breaking is not observed for all the modes in \PCP, suggesting a small amplitude of the structural changes. Furthermore, temperature dependence of parameters for  \textbf{Ph3}, \textbf{Ph4}, and \textbf{Ph6} phonons indicate that some structural changes already occur a higher temperature scale of $\sim$120 K, indicating slow structural changes above the 70 K transition.

\subsection{Crystal Field Excitations of Pr$^{3+}$}

\begin{figure}[h]
    \centering
    \includegraphics[width=\linewidth]{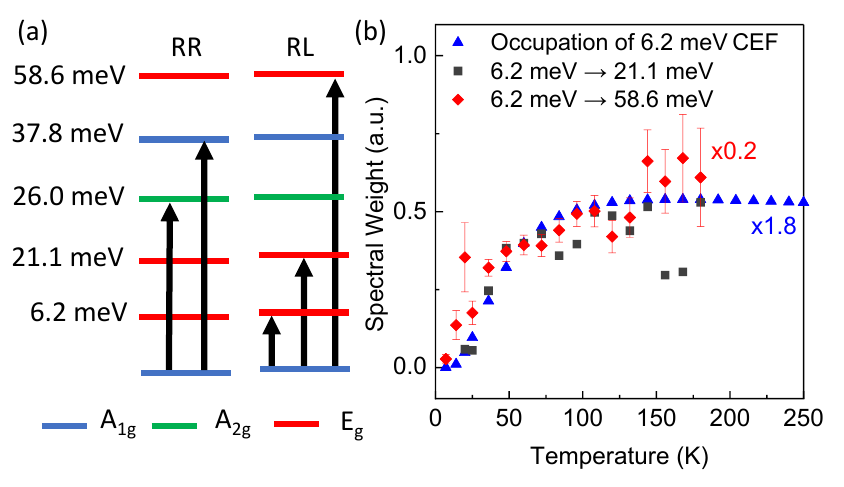}
    \caption{(a) Symmetry-allowed excitations of Pr$^{3+}$ in a $D_{3d}$ crystal field environment in RR and RL polarization channels. Energies of crystal electric field  levels are determined from Raman scattering spectra. (b) Spectral weight of crystal electric field  excitations from the first excited state to the 21.1 and 58.6 meV levels, normalized to spectral weight of \textbf{Ph4} as a proxy for total spectral intensity. Triangles mark temperature dependence of the thermal population of   crystal electric field level at 6.2 meV.}
    \label{fig:interband}
\end{figure}

A comparison of polarized Raman scattering spectra of \PCP\ to \NCP, and strong temperature dependence of the linewidth of some excitations, allows us to assign seven  CEF excitations, as summarized in Table~\ref{tab:CEFcalc_table} and marked with green dashed lines in Fig. \ref{fig:Fig_warming_Raman}. The intensity of the excitations found at 14.8 and 52.1 meV in RR (Fig. \ref{fig:Fig_warming_Raman}b)  decreases below $\sim$60 K, 
and is found proportional to the occupation of the first excited CEF level at 6.2 meV (see Fig. \ref{fig:interband}b). Based on this temperature dependence of intensities and the energies of the excitations, 
we assign these excitations to  the transitions from the first excited state at 6.2 meV to the higher crystal field levels (see Fig. \ref{fig:interband}a and Table~\ref{tab:CEFcalc_table}). The remaining five excitations are assigned to CEF transitions from the ground state.

\begin{table}[h]
    \centering
    \begin{tabular}{|c c c c c c|}
        \hline
        \multicolumn{3}{|c}{CEF energy (meV)} & Polarization & Symmetry & Comments\\
        calc. & Raman & INS$^1$ & & & \\
        \hline
        0 & --- & --- & --- & A$_{1g}$ & \\
        6.5 & 6.2 & 6.5 & XX,XY,RL & E$_{g}$ & \\
        --- & 14.8 & 14.4 & XX,XY,RR & --- & 6.2 $\rightarrow$ 20.4\\
        21.1 & 20.4/21.1 & 20.4 & XX,XY,RL & E$_{g}$ &\\
        26.1 & 26.0 & 26.6 & XY,RR & A$_{2g}$ &\\
        32.9 & 38.8 & 33.2 & XX,RR & A$_{1g}$ &\\
        --- & 52.1 & 52.1 & XX,XY,RR & --- & 6.2 $\rightarrow$ 58.6 \\
        59.1 & 58.6/59.1 & --- & XX,XY,RL & E$_{g}$ &\\
        \hline
\end{tabular}
    \caption{Energies of crystal field excitations of Pr$^{3+}$ in \PCP,  calculated,  observed in Raman scattering spectra, and in inelatic neutron scattering (INS) in Ref.~\cite{GomezAlvarado2025}. Excitations to doubly degenerate (E$_g$) levels are split at low temperatures, which is indicated by the two frequencies listed. The last column shows symmetry of the final state deduced from the polarization dependence of the spectra, with a ground state of A$_{1g}$ symmetry. 
    }
    \label{tab:CEFcalc_table}
\end{table}


In the $D_{6h}$ symmetry of the unit cell, which preserves $D_{3d}$ symmetry of the Pr$^{3+}$ environment and corresponds to 2A$_{1g}$ + 1A$_{2g}$+ 3E$_{g}$ crystal field splitting, we would expect five Raman transitions, where A$_{1g} \rightarrow$ E$_g$ transitions are active in XX, XY, and RL; A$_{1g} \rightarrow$  A$_{1g}$ transitions are active in XX and RR; and  A$_{1g} \rightarrow$  A$_{2g}$ transitions are active in XY and RR scattering channels (see Fig.~\ref{fig:interband}a).  If the symmetry of the lattice is reduced to $D_{2h}$ at low temperatures, we would expect a larger number of CEF excitations, since the degeneracy of all CEF levels will be lifted.  

Point charge model calculations predict a singlet ground state, in contrast to the non-Kramers doublet often observed in Pr$^{3+}$ in a $D_{3d}$ environment \cite{GomezAlvarado2025, Xu2021}. 
Additionally, a plateau observed in magnetic susceptibility below $\sim$25 K indicates a singlet ground state \cite{Jackson2023}. Our experimental spectra are understood well within $D_{3d}$  symmetry if we assume an A$_{1g}$ ground state, with the expected five Raman-active CEF transitions from the ground state, see  Fig.~\ref{fig:interband}a and Table~\ref{tab:CEFcalc_table}.  

Raman CEF frequencies (Table~\ref{tab:CEFcalc_table}) are in agreement with the ones observed in  inelastic neutron scattering (INS)~\cite{GomezAlvarado2025}, except for the CEF at 38.8 meV: INS instead suggests one at 33 meV, which is very weak if distinguishable at all in INS spectra.  Fits of the Stevens operator coefficients reproduce these energies and predict a doublet at 59.1 meV~\cite{GomezAlvarado2025}, which is above the spectral range measured in INS, but is observed by Raman scattering at 58.6/59.1 meV.

The low temperature symmetry breaking, which can be regarded as a small distortion of the D$_{3d}$ site symmetry,  leads to   a splitting of 0.7 meV in the second doublet at 20.4/21.1 meV and a splitting of 0.5 meV in the third doublet at 58.6/59.1 meV.
A splitting of the first excited state doublet observed by INS~\cite{GomezAlvarado2025} is smaller than the spectral resolution of our measurements, which could explain that it is not detected in our Raman spectra.

\section{Discussion}

To summarize, our Raman scattering study demonstrates the low temperature symmetry breaking in \PCP. One of the main results  is the detection of the soft mode, which provides evidence for the displacive structural instability in the  Cd$_{trig}$/P$_{trig}$ hexagonal layer. The structural instability of this layer can also  be a reason for the low frequency shift of experimental frequencies of the Raman active  phonons as compared to the calculated  ones, observed only for  the atoms in this layer (see Table~\ref{tab:phonon_table}).   Our  results are in agreement with the XRD  studies of \PCP\ at 80~K  which  have found an evidence of bond/lattice instability of the  Cd$_{trig}$/P$_{trig}$ honeycomb lattice, when  short-range-ordered CdP pseudo-dimers emerge upon cooling \cite{GomezAlvarado2025}.  This structural change is consistent with the calculated eigenvector of the soft mode.  Such short-range order rather than a phase transition is suggested to be a result of frustration in the choice of dimerized bonds among three equivalent directions.
XRD detects the symmetry breaking at slightly higher temperatures than 70~K suggested by the soft mode behavior.  This indicates  a gradual formation of bond order  with the time scale of the freezing different as probed on different time scales. In fact, the unconventional behavior of phonons \textbf{Ph3}, \textbf{Ph4}, and \textbf{Ph6} starting at $\sim$120 K indicates  gradual formation of bond order, possibly a dynamic one, above the transition.

The symmetry breaking suggested by XRD~\cite{GomezAlvarado2025} and confirmed by our Raman scattering study corresponds to the change of the point group symmetry of the lattice from D$_{6h}$ to D$_{2h}$. This  should lead to the following splitting of phonons and separation of polarizations:   E$_g$ (RL) $\rightarrow$  A$_{1g}$ (RR+RL) + B$_{1g}$ (RL). Indeed, both E$_{2g}$ phonons of  the atoms in the honeycomb Cd$_{trig}$/P$_{trig}$ layer (\textbf{Ph1} and \textbf{Ph6}), as well the \textbf{Ph5} phonon of the Pr$^{3+}$ environment, show large linewidth at low temperatures, suggesting splitting which  is not fully resolved. 
However, we do not observe a well-defined polarization separation between the channels in the broken symmetry state, which  can be a consequence of the presence of small-scale differently oriented domains and some level of structural disorder associated with it. 
Indeed, the XRD study of Ref.~\cite{GomezAlvarado2025} suggests the presence of structural domains (correlation length $\sim$20 \r{A}). 
Interestingly,  non-degenerate phonons stay narrow at low temperatures, demonstrating that the disorder is only related to the symmetry breaking process, and the domains provide sufficient coherence length. 

Out-of-plane phonons show additional sensitivity to the changes in the    interstitial layer:  \textbf{Ph3} (A$_{1g}$), the out-of-plane motion of  Cd$_{tet}$ which connects the two primary structural layers, only becomes present as a coherent excitation  at temperatures below about 70~K. Such sensitivity of out-of-plane phonons to the in-plane symmetry breaking is typical in 2D systems, and in different circumstances results in these phonons being probes of the number of layers of 2D materials during exfoliation ~\cite{Tonndorf2013}. 
 
 \PCP\ behaves as a system of   weakly coupled two dimensional layers, where a displacive structural instability in the Cd$_{trig}$/P$_{trig}$ layer drives a small-amplitude structural distortion in the rare earth layer. The symmetry lowering in the rare earth layer
~is best characterized by the small-amplitude splitting of degenerate CEF levels.

The point charge model is known not to reproduce energies of CEF in \PCP~\cite{GomezAlvarado2025}, which can be  a natural consequence of the changes in the hexagonal layer influencing the charge distribution expected from simple approximations. We can suggest that the low temperature structure of the Cd$_{trig}$/P$_{trig}$ layer produces a potential which can influence \textit{Ln}$^{3+}$ CEF energies and  has a potential to affect  magnetic properties. Potentially, the $\textit{Ln}\textit{M}_3\textit{Pn}_3$ family can provide multiferroic properties which can control magnetism: The structural change in the Cd$_{trig}$/P$_{trig}$ hexagonal layer leads to a formation of Cd-P dimers in the $(ab)$ plane, which could produce electric polarization if they would show directional long range order.  One can speculate that an application of strain would produce a single domain in the Cd$_{trig}$/P$_{trig}$ hexagonal layer, which can carry an in-plane dipole moment, and would strongly affect or even control the magnetism associated with \textit{Ln}$^{3+}$. 

In the $\textit{Ln}\textit{M}_3\textit{Pn}_3$ family, the unit cell size  increases on going from PrCd$_3$P$_3$  to  LaCd$_3$P$_3$  and CeCd$_3$P$_3$~\cite{Jang2025}, which can be thought of as ``negative chemical pressure'' modeling the effect of external hydrostatic pressure. A recent study of the optical conductivity of CeCd$_3$P$_3$ and LaCd$_3$P$_3$
~found  a large amplitude splitting of the doubly-degenerate $E_{1u}$ infrared-active phonons~\cite{Jang2025}. This suggests that the amplitude of the the low-temperature symmetry breaking associated with structural instability in the Cd$_{trig}$/P$_{trig}$ hexagonal layer can be controlled by pressure.


Ref. \cite{Jang2025} suggests a link between the position of the Cd$_{tet}$ and P$_{oct}$ atoms along the c-axis and the DFT-calculated semiconducting gap. According to these calculations, the positions of these atoms corresponding to the structure of \PCP\ would lead to a gap of $>$0.5 meV, consistent with the gap of 0.76 meV calculated by our DFT and the increase in Raman intensity observed on cooling. The   dynamic displacement of these atoms corresponds to the eigenvector of  \textbf{Ph3} phonon (see Fig. \ref{fig:phonons}), which is resolved as a coherent excitation only below temperatures near the structural transition, and broadens at low temperatures, as well as phonon \textbf{Ph4}, which softens slightly at low temperatures. This unconventional behavior may indicate an additional structural instability that could couple to the electronic properties of the compounds in this family.

\section{Conclusion}

In conclusion, temperature-dependent Raman scattering spectroscopy of \PCP\ demonstrates that the structural instability in the hexagonal Cd$_{trig}$/P$_{trig}$ layer is associated with a soft mode, the eigenvector of which  corresponds to atomic displacements observed in XRD. We see minimal splitting of doubly degenerate phonon modes at low temperatures, suggesting that the symmetry-breaking displacements are weak. We assign CEF excitations and symmetries of Pr$^{3+}$ using polarized Raman spectra, confirming a singlet ground state. 
We observe splitting of $>$0.5 meV in the higher energy doublet CEF, confirming that the local environment of the \textit{Ln}$^{3+}$ site (and hence the magnetism) is coupled to the interstitial layer which shows the displacive transition. According to Ref.~\cite{GomezAlvarado2025}, this transition leads to a formation of Cd-P pairs, which can carry an electric dipole moment, and are a potential source of electric polarization in case of a full long-range order. We speculate that materials in this family could become multiferroic under strain that would relieve the frustration in the interstitial layer.  


\section{Acknowledgments}

Research at JHU has been performed at the United States Department of Defense funded Center of Excellence for Advanced Electro-photonics with 2D materials—Morgan State University, under Grant No. W911NF2120213. DR, SJGA, and SDW acknowledge US Department of Energy (DOE), Office of Basic Energy Sciences, Division of Materials Sciences and Engineering under Grant No. DE-SC0017752

\bibliography{PrCd3P3}

\end{document}


\newcommand{\PCP}{PrCd$_3$P$_3$}
\newcommand{\NCP}{NdCd$_3$P$_3$}
\newcommand{\redcite}{\textcolor{red}{(CITE)}}
\newcommand{\wn}{cm$^{-1}$}

\title{Supplementary Information for Raman scattering spectroscopic observation of a ferroelastic crossover in bond-frustrated \PCP}
\author{Jackson Davis}
\affiliation{Department of Physics and Astronomy, Johns Hopkins University, Baltimore, Maryland 21218, USA}
\author{Jesse Liebman}
\affiliation{Department of Physics and Astronomy, Johns Hopkins University, Baltimore, Maryland 21218, USA}
\author{Dibyata Rout}
\affiliation{Materials Department, University of California, Santa Barbara, California 93106, USA}
\author{S.J. Gomez Alvarado}
\affiliation{Materials Department, University of California, Santa Barbara, California 93106, USA}
\author{Stephen D. Wilson}
\affiliation{Materials Department, University of California, Santa Barbara, California 93106, USA}
\author{Natalia Drichko}
\affiliation{Department of Physics and Astronomy, Johns Hopkins University, Baltimore, Maryland 21218, USA}
\date{\today}

\maketitle

\begin{figure}
    \centering
    \includegraphics[width=0.9\linewidth]{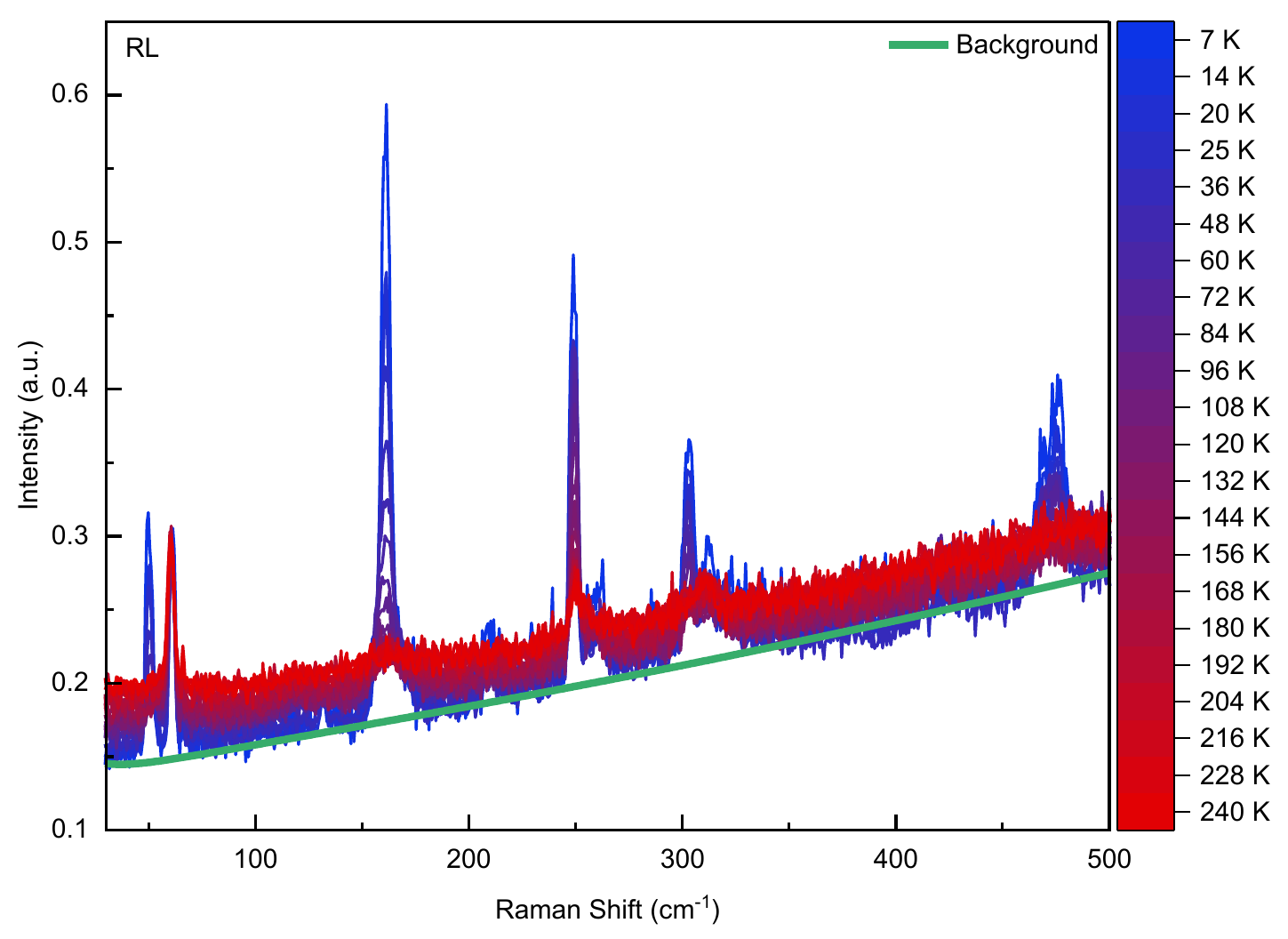}
    \caption{Raman scattering spectra of \PCP$ $ in RL polarization from 7 K to 240 K before subtraction of the temperature-independent background (green).}
    \label{fig:Pr_bgd_RL}
\end{figure}

\section{Background Subtraction}

\begin{figure}
    \centering
    \includegraphics[width=\linewidth]{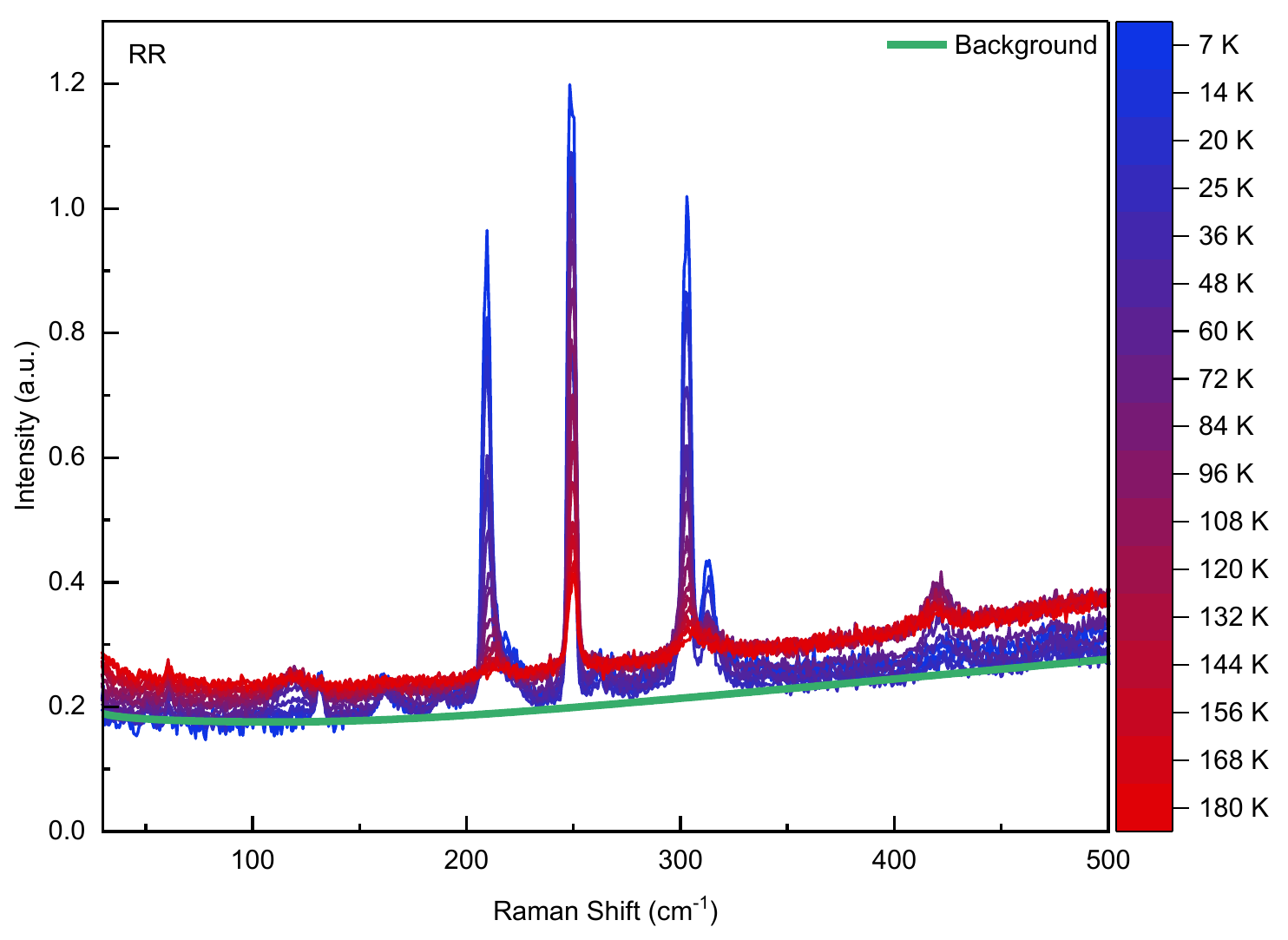}
    \caption{Raman scattering spectra of \PCP$ $ in RR polarization from 7 K to 180 K before subtraction of the temperature-independent background (green).}
    \label{fig:Pr_bgd_RR}
\end{figure}

A temperature-independent background was subtracted from all Raman spectra to highlight the behavior of narrow phonon and CEF features, as described in the main text. The original spectra before background subtraction are presented for RL and RR scattering channels in Figs. \ref{fig:Pr_bgd_RL} and \ref{fig:Pr_bgd_RR} respectively along with the background that was subtracted from all spectra in each channel.

